\documentstyle[12pt,axodraw]{article}
\newcommand{\be}{\begin{equation}}
\newcommand{\ee}{\end{equation}}
\newcommand{\ba}{\begin{eqnarray}}
\newcommand{\ea}{\end{eqnarray}}
\newcommand{\la}{\langle}
\newcommand{\ra}{\rangle}
\begin{document}

\begin{center}
{\large{\bf Susceptibilities to
order $\alpha_s$ in the high density phase of QCD}}\\
\vspace{1cm}
Krishnendu Mukherjee\\
\vspace{0.5cm}
Department of Physics,
Rampurhat College, P.O.- Rampurhat, Dist.- Birbhum,\\
Pin-731224, West Bengal, India.


\vspace{1.5cm}

{\bf Abstract}
\end{center}

We compute the free energy density of QCD to order $\alpha_s$ 
at very high density and non-zero quark masses. The counterterms 
needed to renormalise the theory to order $\alpha_s$ is same as
the vacuum (non-zero density) theory. We investigate the response
of the theory to non-zero quark masses and chemical potentials. We
study quark number density and quark number susceptibility in the high density 
limit, where the ratio of the quark mass to the corresponding 
chemical potential is very small ($m/\mu\ll 1$). In this limit
both number density and susceptibility contain a $\ln(m/\mu)$ 
contribution at order $\alpha_s$. We compute the scalar and 
pseudoscalar susceptibilities to order $\alpha_s$ in three flavour
QCD at high density taking quark masses and chemical potentials 
to be degenerate and non-zero. At extremely high density, 
since $\alpha_s$ is very small, both the susceptibilities are found 
to be same in the chiral limit. This means that the 
scalar-pseudoscalar splitting is absent in the CFL phase.

\vspace{0.5cm}

%
%

\section{Introduction}
\setcounter{equation}{0}
\renewcommand{\theequation}{1.\arabic{equation}}

The study of QCD at very high densities and low temperatures 
has been received a lot of interest in the last few years. It 
has been argued that at a sufficiently high density the
colour symmetry is spontaneously broken, leading to colour
superconductivity \cite{Barrois, Bailin, Alford1, Alford2} (see
\cite{Rajagopal} for recent review). 
This novel phase might be accessible in
the core of neutron stars or in the baryon-rich fragmentation
region of heavy-ion collisions. 

At extremely high density, where the density is of the order 
of 3 to 5 times the saturation density of nuclear matter, the
wave function of the quarks in nucleons will overlap with that
of quarks in other nucleons due to asymptotic freedom.
At such high density quarks are no longer confined in nucleons
and thus the nuclear matter will become a quark matter \cite{Collins}.
The matter at such high density will consist of a Fermi sea of
essentially free quarks and its behaviour is dominated by the 
high momentum quarks that live at the Fermi surface. It was shown 
by Bardeen Cooper and Schriffer (BCS) \cite{BCS} that in presence
of attractive interactions a Fermi surface is unstable. Since
in QCD the attractive interaction is provided by one-gluon 
exchange between quarks in a colour antisymmetric channel, the
true ground state of the system is not the naked Fermi surface,
but rather a complicated coherent state of particle and hole pairs
or condensates, called "Cooper pairs". The formation of condensates
leads to energy gaps for both the (quasi) quarks and some or all
gluons. At asymptotically high density, and because of asymptotic
freedom of QCD, it is possible to compute the quark gap and 
the gluon masses from first principles \cite{Son, Schaefer, Pisarski,
Hong, Brown, Manuel}.

In order to obtain the bulk quantities of a system, such as number
density of particles, susceptibilities etc., it is necessary to obtain
the free energy density or effective potential of the system.
These bulk quantities give us informations about the
behaviour of the system in the microscopic level. Freedman
and McLerran computed the renormalisation group (RG) improved 
thermodynamic potential at finite
density to order $\alpha_s^2$ in non-abelian gauge theory with 
massless quraks \cite{Freedman}. Using this RG improved thermodynamic
potential they obtained the effective quark number density. They
also obtained using RG equation (Gell-Mann-Low equation) 
the flow of strong coupling constant
as a function of chemical potential. Their evaluation shows that
$\alpha_s$ decreases with the increase in density. The two-loop
exchange contribution in the thermodynamic potential for massive
quarks, later, has been evaluated in the ref.\cite{Baym}

In the present work we have studied the susceptibilities to leading
order in strong coupling constant (order $\alpha_s$) in the high 
density phase of QCD. As a first step of this we have computed 
the free energy density to order $\alpha_s$ at finite density taking
quark masses to be non-zero. For massive quarks the cancellation of
ultraviolet (UV) divergences is a bit involved issue. The free energy at
order $\alpha_s$ is UV divergent. This UV divergent piece has two parts
: One is the vacuum (zero density) part and the other one is non-zero
density part. However, both the parts get cancelled from the counterterm
diagrams. The counterterms needed to make the free energy density finite
are the same as what we had in the vacuum theory. 
Using this renormalized effective action we have computed the 
quark number density and the quark number susceptibility to order
$\alpha_s$. We find that at finite baryon density, which is high 
enough too, the density of states of a particular quark flavour 
near the Fermi surface decreases with the increase in mass of the quark.
Therefore, at finite baryon chemical potential the density of states
of strange quark is much lesser than that of $u$ and $d$ quarks near
the Fermi surface. 

The colour and flavour symmetry of QCD with three massless quark 
flavours break at high density. The breaking of this flavour and
colour symmetry leads to eight massive gluons and gapped 
(quasi) quarks. 
The nine (quasi) quarks (three colours times three flavours) 
fall into an ${\bf 8}\oplus{\bf 1}$ of unbroken global SU(3). 
Therefore, in order to parametrise this phase completely 
one needs two gap parameters. 
This particular phase is called the colour-flavour
locked (CFL) phase\cite{Alford2}. In the CFL phase both colour and 
flavour symmetries are dynamically broken by the formation of
quark-quark condensate near the Fermi surface. Since the chiral
symmetry in this phase is broken as well by the formation of
condensates between quark and antiquark pair, the phase can be 
best described by an effective theory of pseudo-Goldstone 
bosons \cite{Casalbuoni, Bedaque}. Since the scalar and pseudoscalar
susceptibilities are related to the order parameter of the chiral
symmetry breaking ($\la\bar{\psi}\psi\ra$, where $\psi$ represents
the quark field) \cite{Kocic}, it is pertinent to 
study both the susceptibilities in the CFL phase.
Using the renormalized free energy density we have computed the 
scalar and pseudoscalar susceptibilities to order $\alpha_s$ in the
high density phase of QCD with three flavours, 
taking quark masses and chemical potentials
to be degenerate and non-zero. We find that at extremely high
density, since $\alpha_s$ is very small, both the susceptibilities are
same.

We organize the paper as follows. In sec.2 we compute the
renormalized free energy density to order $\alpha_s$. 
In sec.3 we compute the susceptibilities. In sec.4 we discuss 
the results. In Appendix-I we give the fermion propagator 
in the dense medium. In Appendix-II, III and IV we discuss the
estimation of integrals which are needed to compute the
free energy density in sec.2.

\section{Computation of free energy density to order $\alpha_s$}
\setcounter{equation}{0}
\renewcommand{\theequation}{2.\arabic{equation}}

The gauge Fixed QCD Lagrangian with $n_f$ number of quark flavours at
finite density is given as
\ba
{\cal L} &=& \bar{\psi}_A(i\partial\!\!\!/ + g A\!\!\!/ 
- m_A + \mu_A\gamma^0)\psi_A - {1\over 4} F^a_{\mu\nu}F^{a\mu\nu}
- {1\over 2\xi}(\partial^\mu A^a_\mu)^2\nonumber\\ 
& &+ \partial_\mu c_a^\dagger (\delta^{ab}\partial^\mu 
- g f^{abc} A^{c\mu}) c_b,
\ea
where $F^a_{\mu\nu} = \partial_\mu A_\nu^a - \partial_\nu A^a_\mu 
- g f^{abc} A_\mu^a A_\nu^b$. $a, b, c$ ($= 1, \cdots 8$) 
are the gauge group indices and $A, B, C$ ($= 1, \cdots n_f$)
are the flavour indices of quarks. $\xi$ is the gauge fixing 
parameter and we shall take $\xi=1$ (Feynman gauge) in the 
following calculations.

\subsection{Counterterms to order $g^2$}

At order $g^2$ the QCD Lagrangian is divergent. The divergences are
coming from the one-loop momentum integrals in the quark and gluon 
two point functions. In order to remove the divergences one needs 
counterterms. The renormalized QCD Lagrangian reads
\be
{\cal L}_{ren} = {\cal L} + {\cal L}_{ct}
\ee
In order to obtain the counterterm Lagrangian let us first 
calculate the quark two point function at order $g^2$ 
(the quark self energy diagram is shown in Fig.1a).
We calculate the self energy diagram in Fig.1a using 
the quark propagator in Appendix-I. This diagram is 
the sum of two parts: The vacuum part and the finite density
part. Since the vacuum part is ultraviolet (UV) finite, we 
shall calculate only the vacuum part of the quark self energy. 
The vacuum part of the self energy reads 
\be
\Sigma^{(vac)}(p)_{AB} = -g^2 C_2(R)\delta_{AB}
\Lambda^{2\epsilon}\int{d^dk\over (2\pi)^d} 
[\gamma^\mu iS(\bar{p} - k)\gamma^\nu] iD_{\mu\nu}(k),
\ee
where the gluon propagator $iD_{\mu\nu}(k) = -{i\eta_{\mu\nu}\over
k^2+i\epsilon}$ (in Feynman gauge), $\Lambda$ is the regularisation scale
and $d$ ($= 4-2\epsilon$, $\epsilon > 0$) is the space-time dimension. Here 
$C_2(R) = {4\over 3}$ (for SU(3)) and $\bar{p} = (p_0+\mu_A, \vec{p})$. 
Performing the $k$ integration in $d$ dimension we obtain
\be
\Sigma^{(vac)}(p)_{AB}(p) = ig^2\delta_{AB} (-1)^{-\epsilon}
{\Gamma(\epsilon)\over 6\pi^2(4\pi\Lambda^2)^{-\epsilon}}
\int_0^1 dt (1-t)^{-\epsilon}t^{-\epsilon}
{(1-\epsilon)t\bar{p}\!\!\!/ - (2-\epsilon)m_A\over
(\bar{p}^2 - {m_A^2\over t})^\epsilon}.
\ee
This term is divergent when $\epsilon\rightarrow 0$. In order 
to remove this divergence we add to ${\cal L}$ the following 
counterterm (the corresponding counterterm diagram is shown
in Fig.1b)
\be
{\cal L}_{ct}^{(1)} = -{g^2\over 6\pi^2(4\pi\Lambda^2)^{-\epsilon}}
(-1)^{-\epsilon}\Gamma(\epsilon)\int_0^1 dt t^{-\epsilon}
(1-t)^{-\epsilon}\bar{\psi}_A{\cal O}\bar{\psi},
\ee
where
\be
{\cal O} = {(1-\epsilon)t\{\gamma^0(i\partial_0 + \mu_A) 
- i\vec{\gamma}.\vec{\nabla}\} - m_A(2-\epsilon)
\over \left[(i\partial_0 + \mu_A)^2 + \nabla^2 - {m_A^2\over t}
\right]^\epsilon}.
\ee
For minimal subtraction we shall retain terms up to zeroth order
in $\epsilon$.

Next we calculate the gluon two point function at order $g^2$
(the gluon polarisation diagram is shown in Fig.2a). 
Although the diagrams with gluons and the ghosts inside the loops
give a non-zero contribution to the gluon polarisation tensor 
at order $g^2$, they contribute zero to the free energy. So
for our purpose we calculate the gluon polarisation taking 
only quarks in the loop. 
The vacuum part of the gluon polarisation (since the finite 
density part is UV finite) reads
\be
\Pi_{\mu\nu}^{(vac)ab}(p) = {g^2\over 2}\delta^{ab}\Lambda^{2\epsilon}
\int{d^dk\over (2\pi)^d}tr_s[iS(k)\gamma^\nu iS(k-p)\gamma^\mu],
\ee
where $tr_s$ means the trace of the product of Dirac gamma-matrices.
We evaluate the traces in $d$ space-time dimension using the
formulae $\gamma^\mu\gamma_\mu = d$ and $tr_s(\gamma_\mu\gamma_\nu)
= f(d)\eta_{\mu\nu}$ ($f(d)\rightarrow 4$ as $d\rightarrow 4$).
We drop the terms linear in $k$ and obtain
\ba
\Pi_{\mu\nu}^{(vac)ab}(p) &=& - f(d)\int_0^1 dt\Lambda^{2\epsilon}
{d^dk\over (2\pi)^d}\nonumber\\
& &\times {2k_\mu k_\nu - \eta_{\mu\nu}k^2
+ \eta_{\mu\nu}\{m_A^2+p^2t(1-t)\} - 2p_\mu p_\nu t(1-t)\over
[k^2 - \{m_A^2 - p^2t(1-t)\}]^2},
\ea
where we have used the Feynman parametrisation.  
Performing the $k$-integration in $d$ dimension we obtain
\ba
\Pi_{\mu\nu}^{(vac)ab}(p) &=& -ig^2f(d)\delta^{ab}
(-1)^{-\epsilon} {\Gamma(\epsilon)\over 
16\pi^2(4\pi\Lambda^2)^{-\epsilon}}
\int_0^1 dt t^{-\epsilon}(1-t)^{-\epsilon}\nonumber\\
& &\times\left[{t(1-t)\eta_{\mu\nu}\over
\left[p^2 - {m_A^2\over t(1-t)}\right]^{-1+\epsilon}}
+ {m_A^2\eta_{\mu\nu} - p_\mu p_\nu t(1-t)\over 
\left[p^2 - {m_A^2\over t(1-t)}\right]^{\epsilon}}\right].
\ea
This term is divergent as $\epsilon\rightarrow 0$. 
In order to remove this divergence we add the 
following counter term to the Lagrangian (the corresponding
counterterm diagram is shown in Fig.2b).
\be
{\cal L}^{(2)}_{ct} = -{g^2\over 32\pi^2(4\pi\Lambda^2)^{-\epsilon}}
f(d)(-1)^{-\epsilon}\Gamma(\epsilon)
\int_0^1 dt t^{-\epsilon}(1-t)^{-\epsilon}A^a_\mu{\cal O}^{\mu\nu}
A^a_{\nu},
\ee
where
\be
{\cal O}^{\mu\nu} = {\eta^{\mu\nu}t(1-t)\over 
\left[-\Box - {m_A^2\over t(1-t)}\right]^{-1+\epsilon}}
+ {m_A^2\eta^{\mu\nu} + t(1-t)\partial^\mu\partial^\nu\over
\left[-\Box - {m_A^2\over t(1-t)}\right]^\epsilon}.
\ee
For minimal subtraction we shall retain terms in ${\cal L}^{(2)}_{ct}$
upto zeroth order in $\epsilon$.
Therefore the renormalized Lagrangian to order $g^2$ is
\be
{\cal L}_{ren} = {\cal L} + {\cal L}^{(0)}_{ct} 
+ {\cal L}^{(1)}_{ct} + {\cal L}^{(2)}_{ct},
\ee
where ${\cal L}^{(0)}_{ct}$ is a field independent counterterm which
is required to renormalise the free ($g$-independent) part of the
free energy density. This can be written as
\be 
{\cal L}^{(0)}_{ct} = {\cal C}(m_A, \epsilon),
\ee
where ${\cal C}$ is a functions of quark masses ($m_A$) and 
$\epsilon$. For minimal subtraction we shall retain in 
${\cal C}$ the terms upto first order in ${1\over \epsilon}$.

\subsection{Free energy density to order $\alpha_s$}

The renormalized QCD partition function reads
\be
Z_{ren} = \int{\cal D}\phi e^{iS_{ren}(\phi)},
\label{ren Z}
\ee
where the variable set $\phi\equiv \{\bar{\psi}, \psi, A_\mu^a, c_a^\dagger,
c_a\}$. The renormalized free energy density is 
\be
F = {1\over i\Omega} \ln Z_{ren},
\label{def of ren F}
\ee
where $\Omega$ is the 4-space-time volume.
Retaining only the $g$-independent quadratic terms in 
$S_{ren}$ we expand the remainder up to order $g^2$ in perturbation 
theory and integrate over $\phi$. The graphs that will contribute
to order $g^2$ are shown in Fig.3. Therefore the renormalized
partition function to order $g^2$ reads
\be
Z_{ren} = Z_{ren}^{(0)}\left[1 + \Omega\sum_{A=1}^{n_f}\left\{
{1\over 2}I_A^{(3{\rm a})}
+ {1\over 2}I_A^{(3{\rm b})}
+ {1\over 2}I_A^{(3{\rm c})}
+ I_A^{(3{\rm d})}
+ I_A^{(3{\rm e})}
+ I_A^{(3{\rm f})}
\right\}\right]
\label{ren Z1 upto alphas} 
\ee
where $I_A^{(3{\rm j})}$ is the contribution of the diagram in Fig.3j
(${\rm j} = {\rm a}, {\rm b}, {\rm c}, {\rm d}, {\rm e}, {\rm f}$).
Here $Z_{ren}^{(0)}$ reads
\be
Z_{ren}^{(0)} = e^{i\Omega{\cal C}}
\prod_{A=1}^{n_f}\left\{
\det[i\partial\!\!\!/ - m_A + \mu_A\gamma^0 
- \epsilon\gamma^0\partial_0]
\right\}^3,
\label{ren Z zero}
\ee 
where the power 3 in the determinant is for three colours 
associated with each flavour of quarks. 
The evaluation of this determinant is discussed in the Appendix-II.
The final result is
\ba
Z_{ren}^{(0)} &=& 
\exp\Biggl[i\Omega\sum_{A=1}^{n_f}\biggl[-{3m_A^4\over 8\pi^2}
\biggl\{{3\over 2} - \gamma_E 
- \ln\biggl({m_A^2\over 4\pi\Lambda^2}\biggr)\biggr\}\nonumber\\
& &+ {1\over 4\pi^2}\biggl\{\mu_A\sqrt{\mu_A^2 - m_A^2}
\biggl(\mu_A^2 - {5\over 2}m_A^2\biggr)
+ {3\over 2} m_A^4\ln\biggl(
{\mu_A + \sqrt{\mu_A^2 - m_A^2}\over m_A}\biggr)\biggr\}\biggr]
\Biggr],
\label{ren Z0}
\ea
where we have chosen ${\cal C} = {3\over 8\pi^2\epsilon}
\sum_{A=1}^{n_f}m_A^4$ for minimal subtraction.

We write in the
following the contribution of the diagrams in 
Fig.3.
\ba
I_A^{(3{\rm a})} &=&
-4ig^2\int{d^4p\over (2\pi)^4}{d^4q\over (2\pi)^4}
{tr_s[iS(p)\gamma^\mu iS(q)\gamma_\mu]\over (p-q)^2},\nonumber\\
I_A^{(3{\rm b})} &=&
-4ig^2\int{d^4p\over (2\pi)^4}{d^4q\over (2\pi)^4}
{A^{\mu\nu\sigma}A_{\mu\nu\sigma}\over p^2q^2(p+q)^2},\nonumber\\
I_A^{(3{\rm c})} &=&
24ig^2\int{d^4p\over (2\pi)^4}{d^4q\over (2\pi)^4}
{p.q\over p^2q^2(p-q)^2},\nonumber\\
I_A^{(3{\rm d})} &=& 
3ig^2\int{d^4p\over (2\pi)^4}{d^4q\over (2\pi)^4}
{B\over p^2q^2},\nonumber\\
I_A^{(3{\rm e})} &=&
{ig^2(-1)^{-\epsilon}\over 2\pi^2(4\pi\Lambda^2)^{-\epsilon}} 
\Gamma(\epsilon)\int_0^1 dt t^{-\epsilon}(1-t)^{-\epsilon}
\int{d^4k\over (2\pi)^4} tr_s[{\cal O}(k)iS(k)],\nonumber\\
I_A^{(3{\rm f})} &=&
-{ig^2(-1)^{-\epsilon}\over 4\pi^2(4\pi\Lambda^2)^{-\epsilon}} 
f(d)\Gamma(\epsilon)\int_0^1 dt t^{-\epsilon}(1-t)^{-\epsilon}
\int{d^4k\over (2\pi)^4} [{\cal O}^{\mu\nu}(k)iD_{\mu\nu}(k)],
\ea
where 
\ba
A^{\mu\nu\sigma} &=& \eta^{\mu\nu}(p-q)^\sigma
+ \eta^{\nu\sigma}(p+2q)^\mu - \eta^{\sigma\mu}(2p+q)^\nu,\nonumber\\
B &=& (\eta^{\mu\nu}\eta^{\lambda\sigma} 
- \eta^{\mu\sigma}\eta^{\nu\lambda})
(\eta_{\mu\nu}\eta_{\lambda\sigma} 
- \eta_{\mu\sigma}\eta_{\nu\lambda}),\nonumber\\
{\cal O}(k) &=& {(1-\epsilon)tk\!\!\!/ - m_A(2-\epsilon)\over
\left[k^2 - {m_A^2\over t}\right]^\epsilon},\nonumber\\
{\cal O}^{\mu\nu}(k) &=&
{\eta^{\mu\nu} t(1-t)\over 
\left[k^2 - {m^2_A\over t(1-t)}\right]^{-1+\epsilon}}
+ {m_A^2\eta^{\mu\nu} - t(1-t)k^\mu k^\nu\over
\left[k^2 - {m^2_A\over t(1-t)}\right]^\epsilon}.
\ea
Only the diagrams in Fig.3a, 3e and 3f will give non-zero
contribution to the free energy, the rest will not contribute to
the free energy i.e. $I_A^{(3{\rm b})} = I_A^{(3{\rm c})} = 
I_A^{(3{\rm d})} = 0$.
The integral $I_A^{(3{\rm a})}$ can be written as the sum of three terms
: The vacuum part, density dependent UV divergent part and density
dependent finite part. So we write it as
\be
I_A^{(3{\rm a})} = 4ig^2 [f(d) I_A^{(0)} + f(d) I_A^{(1)} + I_A^{(2)}],
\label{Ia}
\ee
where we have evaluated the trace over the Dirac gamma matrices 
in $d$ space-time dimension  and write
the $d$-dimensional integrals in the following:
\ba
I_A^{(0)} &=& \Lambda^{4\epsilon}\int
{d^dk\over (2\pi)^d} {d^dp\over (2\pi)^d}
{(2-d)k.p + m_A^2 d\over (k^2-m_A^2+i\epsilon)(p^2-m^2_A+i\epsilon)
\{(p-k)^2+i\epsilon\}},\nonumber\\
I_A^{(1)} &=& 4\pi i\int {d^4k\over (2\pi)^4}
\delta(k^2-m_A^2)\theta(k_0)\theta(\mu_A - k_0)
\Lambda^{2\epsilon}\int{d^dp\over (2\pi)^d}\nonumber\\
& &\times{(2-d)k.p + m_A^2 d\over (k^2-m_A^2+i\epsilon)(p^2-m^2_A+i\epsilon)
\{(p-k)^2+i\epsilon\}},\nonumber\\
I_A^{(2)} &=& -16\pi^2\int
{d^4k\over (2\pi)^4}{d^4p\over (2\pi)^4}
\delta(k^2-m_A^2)\delta(p^2-m_A^2)\theta(k_0)\theta(p_0)
\theta(\mu_A-k_0)\theta(\mu_A-p_0)\nonumber\\
& &\times{2m_A^2-k.p\over m_A^2-k.p}.
\ea
The integral $I_A^{(3{\rm e})}$ can be splitted up into two parts:
The vacuum part and the density dependent UV divergent part.
Going over to $d$ space-time dimension and evaluating the spinor
traces we can write it as  
\be
I_A^{(3{\rm e})} = {ig^2\over 2\pi^2} f(d) [J_A^{(0)} + J_A^{(1)}],
\label{Ie}
\ee
where
\ba
J_A^{(0)} &=& {i (-1)^{-\epsilon}\over
(4\pi\Lambda^2)^{-\epsilon}}\Gamma(\epsilon)
\int_0^1 dt t^{-\epsilon}(1-t)^{-\epsilon} 
\Lambda^{2\epsilon}\int{d^dk\over (2\pi)^d}
{(1-\epsilon)t k^2 - m_A^2(2-\epsilon)\over
 (k^2 - m_A^2 + i\epsilon)\left[k^2 - {m_A^2\over t}\right]^\epsilon},
\nonumber\\
J_A^{(1)} &=& - m_A^2\left({m_A^2\over 4\pi\Lambda^2}\right)^{-\epsilon}
{\Gamma(\epsilon)\over 1-2\epsilon}\left[{1\over 2-2\epsilon}
-(2-\epsilon)\right]\nonumber\\
& &\times\int{d^4k\over (2\pi)^4} 2\pi\delta(k^2-m_A^2)\theta(k_0)
\theta(\mu_A-k_0) .
\ea
In a similar fashion we can write $I_A^{(3{\rm f})}$, going over to
$d$ space-time dimension as
\be
I_A^{(3{\rm f})} = -{ig^2\over 4\pi^2} f(d) K_A^{(0)},
\label{If}
\ee
where
\be
K_A^{(0)} = -{i(-1)^{-\epsilon}(d-1)\over (4\pi\Lambda^2)^{-\epsilon}}
\Gamma(\epsilon)\int_0^1 dt t^{1-\epsilon} (1-t)^{1-\epsilon}
\Lambda^{2\epsilon}\int{d^dk\over (2\pi)^d}
{1\over \left[k^2 - {m_A^2\over t(1-t)}\right]^\epsilon}.
\ee
In the minimal subtraction scheme we shall retain terms in 
$J_A^{(0)}$, $J_A^{(1)}$ and $K_A^{(0)}$ upto 
first order in ${1\over \epsilon}$.
The evaluation of the integrals $I_A^{(0)}$, $I_A^{(1)}$, $I_A^{(2)}$, 
$J_A^{(0)}$ and 
$J_A^{(1)}$ are given in the Appendix (III and IV). 
The results are written in the following.
\ba
I_A^{(0)} &=& \left({m_A^2\over 16\pi^2}\right)^2
\left({m_A^2\over 4\pi\Lambda^2}\right)^{-2\epsilon}
\left[-{3\over \epsilon^2} - {7\over \epsilon}
+ {6\gamma_E\over \epsilon} - 15 + 14\gamma_E - 6\gamma_E^2 
- {\pi^2\over 2}\right],\nonumber\\
I_A^{(1)} &=& - {m_A^2\over 64\pi^4}
\left[\mu_A\sqrt{\mu_A^2-m_A^2} - m_A^2
\ln\left({\mu_A + \sqrt{\mu_A^2-m_A^2}\over m_A}\right)\right]\nonumber\\
& &\times\left\{{3\over \epsilon}
\left({m_A^2\over 4\pi\Lambda^2}\right)^{-\epsilon}
+ 4 - 3\gamma_E\right\},\nonumber\\
I_A^{(2)} &=& -{1\over 16\pi^4}(\mu_A^2-m_A^2)(\mu_A^2-2m_A^2)
+ {3m_A^2\over 8\pi^4}\mu_A\sqrt{\mu_A^2-m_A^2}
\ln\left( {\mu_A+\sqrt{\mu_A^2-m_A^2}\over m_A}\right)\nonumber\\
& &- {m_A^2\mu_A^2\over 2\pi^4}
\ln\left( {\mu_A+\sqrt{\mu_A^2-m_A^2}\over m_A}\right)
+ {m_A^4\over 16\pi^4}
\ln^2\left( {\mu_A+\sqrt{\mu_A^2-m_A^2}\over m_A}\right)\nonumber\\
& &- {m_A^2(\mu_A-m_A)^2\over 4\pi^4}
\ln\left({\sqrt{\mu_A+m_A} + \sqrt{\mu_A-m_A}
\over 2\sqrt{\mu_A}}\right), \nonumber\\
J_A^{(0)} &=& {m_A^4\over 32\pi^2}\left({m_A^2\over 4\pi\Lambda^2}
\right)^{-2\epsilon}\left[{3\over \epsilon^2} + {7\over \epsilon}
- {6\over \epsilon}\gamma_E\right],\nonumber\\
J_A^{(1)} &=& {3m_A^2\over 16\pi^2\epsilon}
\left({m_A^2\over 4\pi\Lambda^2}\right)^{-\epsilon}
\left[\mu_A\sqrt{\mu_A^2-m_A^2} - m_A^2
\ln\left({\mu_A + \sqrt{\mu_A^2-m_A^2}\over m_A}\right)\right],\nonumber\\
K_A^{(0)} &=& {m_A^4\over 32\pi^2}
\left({m_A^2\over 4\pi\Lambda^2}\right)^{-2\epsilon}
\left[{3\over \epsilon^2} + {7\over \epsilon} 
- {6\over \epsilon}\gamma_E\right].
\label{IJK}
\ea 
The vacuum UV divergences in the diagram Fig.3a get cancelled 
from the counterterm diagrams Fig.3e and 3f and the density
dependent UV divergence in Fig.3a gets cancelled from the 
counterterm diagram Fig.3e. Therefore the renormalized free energy
density to order $\alpha_s$ reads
\be
F = \sum_{A=1}^{n_f} f_A,
\ee
where 
\ba
f_A &=& -{3m_A^4\over 8\pi^2}\Biggl[{3\over 2} - \gamma_E
- \ln\left({m_A^2\over 4\pi\Lambda^2}\right)\Biggr]\nonumber\\
& &+ {1\over 4\pi^2}\Biggl\{\mu_A\sqrt{\mu_A^2-m_A^2}
\left(\mu_A^2 - {5\over 2}m_A^2\right)
+ {3\over 2}m_A^4\ln\left({\mu_A + \sqrt{\mu_A^2-m_A^2}\over m_A}\right)
\Biggr\}\nonumber\\
& &+ {\alpha_s\over \pi^3}\Biggl[
- {m_A^4\over 8}\Bigl(15 - 14\gamma_E + 6\gamma_E^2 + {\pi^2\over 2}
\Bigr) - {1\over 2}(\mu_A^2 - m_A^2)(\mu_A^2-2m_A^2)\nonumber\\
& &- \Bigl(2-{3\over 2}\gamma_E\Bigr)m_A^2\mu_A\sqrt{\mu_A^2-m_A^2}
+ {m_A^4\over 2}\ln^2\left({\mu_A+\sqrt{\mu_A^2-m_A^2}\over m_A}\right)
\nonumber\\
& &+ m_A^2\Bigl\{\Bigl(2-{3\over 2}\gamma_E\Bigr)m_A^2
+ 3\mu_A\sqrt{\mu_A^2-m_A^2} - 4\mu_A^2\Bigr\}
\ln\left({\mu_A+\sqrt{\mu_A^2-m_A^2}\over m_A}\right)\nonumber\\
& &- 2m_A^2(\mu_A-m_A)^2 
\ln\left({\sqrt{\mu_A+m_A} + \sqrt{\mu_A-m_A}
\over 2\sqrt{\mu_A}}\right)\Biggr] + 0(\alpha_s^2).
\ea

\section{Response to chemical potentials and non-zero quark masses}
\setcounter{equation}{0}
\renewcommand{\theequation}{3.\arabic{equation}}

In this section we shall investigate the response of the system 
with the change in chemical potentials and quark masses. The 
quark number density is obtained by taking a derivative of 
the free energy density with respect to chemical potential.
In the high density limit this is given by 
\ba
n_A &=& {\partial F\over \partial\mu_A}\nonumber\\
&=& {\mu_A^3\over \pi^2} - {3\over 2\pi^2}m_A^2\mu_A
+ {\alpha_s\over \pi^3}\biggl[-2\mu_A^3 + (8+3\gamma_E)m_A^2\mu_A
- 4m_A^3 - 14m_A^2\mu_A\ln\left({m_A\over 2\mu_A}\right)\biggr]\nonumber\\
& &+ 0(\alpha_s, {1\over \mu_A}).
\ea
Here we have retained terms upto zero-th order in $\mu_A$. In the chiral
limit with non-interacting quarks the chemical potential 
$\mu_A \sim n_A^{1\over 3}$. However in the non-chiral limit with
interacting quarks the chemical potential cannot take such a simple form 
in terms of density.

The quark number susceptibility to order $\alpha_s$ in the
high density limit (${m_A\over \mu_A}\ll 1$) reads 
\ba
\chi_A &=& {\partial n_A\over \partial\mu_A}\nonumber\\
&=& {3\over \pi^2}\mu_A^2 - {3\over 2\pi^2}m_A^2
+ {\alpha_s\over \pi^3}\biggl[ - 6\mu_A^2 + 3(4+\gamma_E)m_A^2
- 14m_A^2\ln\left({m_A\over 2\mu_A}\right)\biggr]\nonumber\\
& &+ 0(\alpha_s^2, {1\over \mu_A}),
\ea
where we have retained terms upto zero-th order in $\mu_A$.
The susceptibility has a simple physical interpretation. It 
measures the density of states of a particular quark flavour 
near the Fermi surface. At finite baryon density ($\mu_A = \bar{\mu}$
for all $A$, where $\bar{\mu}$ is the baryon chemical potential) 
as the mass of the fermion increases the
susceptibility or the density of states near the Fermi surface 
decreases. Therefore the density of states of the strange quark
near the Fermi surface at finite baryon density is smaller compared 
to the $u$ and $d$ quarks.

The colour and flavour symmetry of QCD with three massless quark 
flavours break at high density. This breaking leads to eight massive
gluons and gapped (quasi) quarks. 
This particular phase is
called CFL phase \cite{Alford2}.
We shall calculate in this phase the scalar and pseudoscalar 
susceptibilities which are related to the chiral order
parameter of the theory. In order to do this we 
consider the three flavour QCD with $u$, $d$ and $s$ quarks
at high density. We take quark masses and chemical potentials 
to be degenerate; $m_u=m_d=m_s =m$ and $\mu_u=\mu_d=\mu_s=\mu$. 
This means that the theory respect the $SU(3)_{L+R}$ flavour symmetry.
Let us define the following scalar and pseudoscalar correlations:
\ba
C_S(q) &=& \int d^4x e^{iq.x}\la 0\mid\bar{\psi}(x)\psi(x)
\bar{\psi}(0)\psi(0)\mid 0\ra,\nonumber\\
C_P(q) &=& \int d^4x e^{iq.x}\la 0\mid\bar{\psi}(x)i\gamma_5\psi(x)
\bar{\psi}(0)i\gamma_5\psi(0)\mid 0\ra.
\ea
The scalar and pseudoscalar susceptibilities are defined as,
\be
\chi_S = C_S(0) ~~~~~ {\rm and} ~~~~~ \chi_P = C_P(0).
\ee
These two susceptibilities are related to the chiral order 
parameter in the following way \cite{Kocic}
\be
\chi_S = {\partial\la\bar{\psi}\psi\ra\over \partial m}~~~~{\rm and}~~~
\chi_P = {\la\bar{\psi}\psi\ra\over m},
\label{def of S and P}
\ee
where $\la\bar{\psi}\psi\ra = \la\bar{u}u + \bar{d}d + \bar{s}s\ra$.
The chiral condensate is obtained by differentiating the free energy 
density with respect to $m$. Once we obtain $\la\bar{\psi}\psi\ra$ 
we can compute the scalar and pseudoscalar susceptibilities using
eqn.(\ref{def of S and P}). Both the scalar and pseudoscalar susceptibilities
in the high density limit read
\ba
\chi_S &=& -{9m^2\over 2\pi^2}\biggl[1-3\gamma_E-3\ln\left({m^2\over 
4\pi\Lambda^2}\right)\biggr]
+ {3\over \pi^2}\biggl[\mu^2 - 3m^2 - {9\over 2}m^2
\ln\left({m\over 2\mu}\right)\biggr]\nonumber\\
& &+ {3\alpha_s\over \pi^3}\Biggl[-(13-3\gamma_E)\mu^2
- \biggl\{19 - {39\over 2}\gamma_E + 9\gamma_E^2 
+ {3\over 4}\pi^2\biggr\}m^2\nonumber\\
& &+ (18\gamma_E -5)m^2\ln\left({m\over 2\mu}\right)
+ 6m^2\ln^2\left({m\over 2\mu}\right)
+ 8\mu^2\ln\left({m\over 2\mu}\right)\Biggr]
+ 0(\alpha_s^2, {1\over \mu}),\nonumber\\
\chi_P &=& -{9m^2\over 2\pi^2}\biggl[1-\gamma_E-\ln\left({m^2\over
4\pi\Lambda^2}\right)\biggr]
+ {3\over \pi^2}\biggl[\mu^2 - {m^2\over 2} - {3\over 2}m^2
\ln\left({m\over 2\mu}\right)\biggr]\nonumber\\
& &+ {3\alpha_s\over \pi^3}\Biggl[(8-3\gamma_E)\mu^2
- \biggl\{10 - {17\over 2}\gamma_E + 3\gamma_E^2 
+ {1\over 4}\pi^2\biggr\}m^2\nonumber\\
& &- (1-6\gamma_E)m^2\ln\left({m\over 2\mu}\right)
+ 2m^2\ln^2\left({m\over 2\mu}\right)
+ 2\mu^2\ln\left({m\over 2\mu}\right)\Biggr]
+ 0(\alpha_s^2, {1\over \mu}).
\ea
At very high density (${m\over \mu}\ll 1$) as we take the chiral limit
($m\rightarrow 0$) both the susceptibilities are logarithmically 
divergent at order $\alpha_s$, i.e. 
$\chi_S\sim{3\over \pi^2}\mu^2\biggl\{1
- {\alpha_s\over \pi}(13-3\gamma_E) + {8\alpha_s\over \pi}
\ln\left({m\over 2\mu}\right)\biggr\}$ and
$\chi_P\sim{3\over \pi^2}\mu^2\biggl\{1
+ {\alpha_s\over \pi}(8-3\gamma_E) + {2\alpha_s\over \pi}
\ln\left({m\over 2\mu}\right)\biggr\}$. 
Since at very high density $\alpha_s\ll 1$ \cite{Freedman}, the resulting 
contribution at order $\alpha_s$ is small compared to the 
leading order (zero-th order in $\alpha_s$) one. Therefore, at 
extremely high density both scalar and pseudoscalar 
susceptibilities are same in the chiral limit, i.e. 
$\chi_S = \chi_P = {3\over \pi^2}\mu^2$.

\section{Discussion and Conclusions}
\setcounter{equation}{0}
\renewcommand{\theequation}{3.\arabic{equation}}

We have computed the free energy density 
of QCD to order $\alpha_s$ at finite density taking 
quark masses to be non-zero. The free energy density is 
found to UV divergent. The counterterms needed to renormalise it
is the same as what we had in the vacuum theory. We have 
computed the quark number density and the quark number susceptibility
in the high density limit ($m_A/\mu_A\ll 1$) using this 
free energy density. In this limit both of them contain a
$\ln(m_A/\mu_A)$ contribution at order $\alpha_s$.

We have computed the scalar and pseudoscalar susceptibility 
in the high density phase of three flavour QCD taking masses and
chemical potentials of the quarks to be degenerate and non-zero.
We find that at very high density, since $\alpha_s\ll 1$, both
the susceptibilities are same in the chiral limit. 
Since the mass is related to the susceptibility 
via $M^2 = Z\chi^{-1}$ \cite{Amit},
where $Z$ is the wavefunction renormalisation, and since the scalars
and pseudoscalars renormalise in the same way, we obtain the following
equality in the chiral limit: $M_S^2 = M_P^2$. Therefore in the 
three flavour QCD both the scalar and pseudoscalar masses are degenerate 
in the CFL phase.

\vspace{0.8cm}

\noindent{\bf Acknowledgements:} I would like to thank Debadesh Bandopadhyay,
Subhasis Basak, Amit Kundu and Pushan Majumdar for useful discussions.

\vspace{0.9cm}

\begin{center}
{\bf Appendix-I: Fermion propagator at finite density} 
\end{center}  
\setcounter{equation}{0}
\renewcommand{\theequation}{I.\arabic{equation}}

Free fermion propagator at finite density reads
\ba
iS(x) &=& \int {d^4p\over (2\pi)^4} e^{-ip.x}
{i\over p\!\!\!/ - m - \mu\gamma^0 + i\epsilon\gamma^0p_0}\nonumber\\
&=& i\int {d^4p\over (2\pi)^4} e^{-ip.x}
{\gamma^0(p_0+\mu) - \vec{\gamma}.\vec{p} + m
\over (p_0 + \mu + i\epsilon p_0)^2 - \omega_p^2},
\ea
where $\omega_p = \sqrt{\vec{p}^2 + m^2}$. Performing the $p_0$
integration we obtain
\ba
iS(x) &=& \theta(x^0) \int {d^3p\over (2\pi)^3}{p\!\!\!/ + m\over 2p_0}
\theta(p_0-\mu) e^{-ip.x + i\mu x^0}\nonumber\\
& & \theta(-x^0) \int {d^3p\over (2\pi)^3}
\biggl[{p\!\!\!/ + m\over 2p_0} 
\theta(\mu - p_0) e^{-ip.x + i\mu x^0}
+ {p\!\!\!/ - m\over 2p_0} e^{ip.x + i\mu x^0}\biggr],
\ea
where $p_0 = \omega_p$. After doing some algebra we 
can write it as 
\be
iS(x) = \int{d^4p\over (2\pi)^4} e^{-ip.x + i\mu x^0} iS(p),
\ee
where $iS(p) = (p\!\!\!/+m)i\tilde{S}(p)$ and 
\be
i\tilde{S}(p) = {i\over p^2 - m^2 + i\epsilon}
- 2\pi\delta(p^2-m^2)\theta(p_0)\theta(\mu - p_0).
\ee

\vspace{0.5cm}

\begin{center}
{\bf Appendix-II: Evaluation of $Z_{ren}^{(0)}$} 
\end{center}  
\setcounter{equation}{0}
\renewcommand{\theequation}{II.\arabic{equation}}

Define 
\be
D=i\partial\!\!\!/ - m_A + \mu_A\gamma^0 
- \epsilon\gamma^0\partial_0.
\ee
Now 
\be
\det{D} = e^{Tr\ln{D}},
\ee
where $Tr$ is taken over spin matrices and space-time.
\be
Tr\ln{D} = \int d^4x \la x\mid tr_s\ln{D}\mid x\ra.
\ee
Inserting complete set of eigenstates of momentum in between 
and using the formulae $\la k\mid x\ra = e^{-ik.x}$,
$\la k^\prime\mid k\ra = (2\pi)^4\delta^{4}(k-k^\prime)$ we
obtain
\be
Tr\ln{D} = \int d^4x {d^4k\over (2\pi)^4} tr_s\ln{D(k)}.
\ee 
Here we have used the fact that operator $D$ is diagonal
in the momentum representation i.e. 
$\la k\mid D\mid k^\prime\ra = (2\pi)^4\delta^{4}(k-k^\prime)D(k)$.
Using the formula $tr\ln{D} = \ln\det{D}$ we obtain
\be
Tr\ln{D} = \int d^4x {d^4k\over (2\pi)^4}\ln{\rm det}_s{D(k)},
\ee 
where the $\det_s$ means the determinant of the resulting 
spin matrix. Evaluating the determinant of the spin matrix we
obtain
\be
Tr\ln{D} = 2\int d^4x {d^4k\over (2\pi)^4}\ln(\bar{k}^2 - m_A^2),
\ee
where $\bar{k} = (k_0 + \mu_A + i\epsilon k_0, \vec{k})$.
Therefore the renormalized free partition function reads,
\ba
Z_{ren}^{(0)} &=& e^{i\Omega{\cal C}}
\prod_{A=1}^{n_f}\left\{\det{D}\right\}^3\nonumber\\
&=& e^{i\Omega{\cal C}} \exp\left\{6\Omega\sum_{A=1}^{n_f}
\int{d^4k\over (2\pi)^4}\ln(\bar{k}^2 - m_A^2)\right\}. 
\ea
We use the following parametrisation 
\be
\ln(\bar{k}^2 - m_A^2) =
\int_0^{\mu_A+\omega_{kA}}{dx\over k_0 + (1- i\epsilon)x}
+ \int_0^{\mu_A-\omega_{kA}}{dx\over k_0 + (1- i\epsilon)x},
\ee
where $\omega_{kA} = \sqrt{\vec{k}^2 + m_A^2}$. Here we have
ignored the mass($m_A$) and density($\mu_A$) independent terms.
We first do the $k_0$-integration going over to the 
$d$ space time dimension and then do the integration over $x$. 
Next we do the remaining $k$ integrations in $d-1$ and 3 dimension
for the vacuum and the finite density part respectively. 
The result is
\ba
\Lambda^{2\epsilon}\int{d^dk\over (2\pi)^d}\ln(\bar{k}^2 - m_A^2)
&=& -{im_A^4\over 16\pi^2}\biggl[{1\over \epsilon}
+ {3\over 2} - \gamma_E - \ln\left({m_A^2\over 4\pi\Lambda^2}\right)
\biggr]
\nonumber\\
& &+ {i\over 24\pi^2}\biggl[\mu_A\sqrt{\mu_A^2 - m_A^2}
\left(\mu_A^2 - {5\over 2}m_A^2\right)\nonumber\\
& &+ {3\over 2}m_A^4
\ln\left({\mu_A + \sqrt{\mu_A^2 - m_A^2}\over m_A}\right)\biggr].
\ea

\vspace{0.5cm}

\begin{center}
{\bf Appendix-III: Evaluation of $I_A^{(0)}$, 
$I_A^{(1)}$ and $I_A^{(2)}$} 
\end{center}  
\setcounter{equation}{0}
\renewcommand{\theequation}{III.\arabic{equation}}

Performing the momentum integration over $k$ and $p$ 
the integral $I_A^{(0)}$ reads
\ba
I^{(0)}_A 
&=& \left({m_A^2\over 16\pi^2}\right)^2
\left({m_A\over 4\pi\Lambda^2}\right)^{-2\epsilon}(2-\epsilon)
\Gamma(-1+2\epsilon)\int_0^1 dy y^{-2+\epsilon}\nonumber\\
& &\times\biggl\{
(2y-1)\int_0^1 dx x^{-\epsilon}(1-xy)^{-2+\epsilon}
+ (1-y)\int_0^1 dx x^{-\epsilon}(1-xy)^{-3+\epsilon}\biggr\}.
\ea
Define 
\be
I_n = \int_0^1 dx x^{-\epsilon}(1-xy)^{-n}~~~ (n\geq 0),
\ee
where it satisfies the following recursion relation
\be
I_n = {1\over 1-\epsilon}[(1-y)^{-n} + n (I_n - I_{n+1})].
\ee   
Using this recursion relation we obtain,
\be
I_{2-\epsilon} = {(1-y)^{-1+\epsilon}\over 1-\epsilon}, ~~~~
I_{3-\epsilon} = {(1-y)^{-2+\epsilon}\over 2-\epsilon} 
+ {(1-y)^{-1+\epsilon}\over (1-\epsilon)(2-\epsilon)}.
\label{Is}
\ee
Therefore we can write 
\ba
I_A^{(0)} =  
&=& \left({m_A^2\over 16\pi^2}\right)^2
\left({m_A\over 4\pi\Lambda^2}\right)^{-2\epsilon}(2-\epsilon)
\Gamma(-1+2\epsilon)\int_0^1 dy y^{-2+\epsilon}\nonumber\\
& &\times\{ (2y-1)I_{2-\epsilon} + (1-y)I_{3-\epsilon}\}.
\ea
Using eqn.({\ref{Is}) and performing the integration over 
$y$ we obtain
\be
I_A^{(0)}  
= \left({m_A^2\over 16\pi^2}\right)^2
\left({m_A\over 4\pi\Lambda^2}\right)^{-2\epsilon}
\left[-{3\over \epsilon^2} - {7\over \epsilon}
+ {6\over \epsilon}\gamma_E - 15 + 14\gamma_E
- 6\gamma_E^2 - {\pi^2\over 2}\right].
\ee

Performing the integration over $p$ the integral $I_A^{(1)}$ reads
\ba
I^{(1)}_A &=& {m_A^2\over 8\pi^2}
\left({m_A^2\over 4\pi\Lambda^2}\right)^{-\epsilon}
\left[-{3\over \epsilon} - 4 + 3\gamma_E\right]\nonumber\\
& &\times\int{d^3k\over (2\pi)^3 2\omega_{kA}}
\theta(\mu_A - \omega_{kA})\nonumber\\
&=& {m_A^2\over 64\pi^4}
\left({m_A^2\over 4\pi\Lambda^2}\right)^{-\epsilon}
\left[-{3\over \epsilon} - 4 + 3\gamma_E\right]\nonumber\\
& &\times\left[\mu_A\sqrt{\mu_A^2 - m_A^2}
- m_A^2\ln\left({\mu_A + \sqrt{\mu_A^2 - m_A^2}\over m_A}\right)\right],
\ea
where  
\be
\int{d^3k\over (2\pi)^3 2\omega_{kA}}
\theta(\mu_A - \omega_{kA})
= {1\over 8\pi^2}
\left[\mu_A\sqrt{\mu_A^2 - m_A^2}
- m_A^2\ln\left({\mu_A + \sqrt{\mu_A^2 - m_A^2}\over m_A}\right)\right].
\label{fd phase space}
\ee

The integral $I_A^{(2)}$ reads
\be
I_A^{(2)} = -{1\over 16\pi^4}\left[\mu_A\sqrt{\mu_A^2 - m_A^2}
- m_A^2\ln\left({\mu_A + \sqrt{\mu_A^2 - m_A^2}\over m_A}\right)\right]
+ {m_A^2\over 8\pi^4}L_A,
\ee
where
\be
L_A = 2\mu_A^2\biggl[(1-\alpha)^2\ln\alpha
+ \int_\alpha^1dx\int_\alpha^1dy
\Bigl\{\ln|x-y| 
- \ln|xy + \sqrt{x^2-\alpha^2}\sqrt{y^2-\alpha^2}-\alpha^2|\Bigr\}\biggr]
\ee
and $\alpha=m_A/\mu_A$. We can carry out the first integral and
the result is
\be
\int_\alpha^1dx\int_\alpha^1dy\ln|x-y|
= (1-\alpha)^2\ln(1-\alpha) 
- {3\over 2}(1-\alpha)^2.
\ee
In order to do the second integral we change the variables from
$x$ and $y$ to $\theta$ and $\phi$ respectively where $x=\alpha\cosh\theta$
and $y=\alpha\cosh\phi$. The integral becomes
\ba
& &\int_\alpha^1dx\int_\alpha^1dy\ln|xy 
-\sqrt{x^2-\alpha^2}\sqrt{y^2-\alpha^2} - \alpha^2|\nonumber\\
&=& 2(1-\alpha)^2\ln\alpha
+ (1-\alpha)^2\ln{2}
+ 2\alpha^2\int_0^ud(\cosh\theta)\int_0^ud(\cosh\phi)\nonumber\\
& &\times\ln\sinh\left({\theta + \phi\over 2}\right),
\ea
where $u = \ln\left({1+\sqrt{1-\alpha^2}\over \alpha}\right)$.
The remaining integration over $\theta$ and $\phi$ can be done
easily. Finally the result for $L_A$ is
\ba
L_A &=& \mu_A^2 - m_A^2 
- 4\mu_A^2\ln\left({\mu_A+\sqrt{\mu_A^2-m_A^2}\over m_A}\right)\nonumber\\
& &+ 2\mu_A\sqrt{\mu_A^2-m_A^2}
\ln\left({\mu_A+\sqrt{\mu_A^2-m_A^2}\over m_A}\right)
+ m_A^2\ln^2\left({\mu_A+\sqrt{\mu_A^2-m_A^2}\over m_A}\right)\nonumber\\
& &- 2(\mu_A-m_A)^2
\ln\left({\sqrt{\mu_A+m_A}+\sqrt{\mu_A-m_A}\over 2\sqrt{\mu_A}}\right).
\ea 

\vspace{0.5cm}

\begin{center}
{\bf Appendix-IV: Evaluation of $J_A^{(0)}$ and $J_A^{(1)}$} 
\end{center} 
\setcounter{equation}{0}
\renewcommand{\theequation}{IV.\arabic{equation}}

Performing the integration over $k$ the integral $J_A^{(0)}$ reads
\be
J_A^{(0)} = -{m_A^4\over 16\pi^2} 
\left({m_A\over 4\pi\Lambda^2}\right)^{-2\epsilon}
\Gamma(-2+\epsilon)(1-\epsilon)(2-\epsilon)[X_1 - X_2],
\ee
where
\ba
X_1 &=& \int_0^1 dy y^{-\epsilon}(1-y)^{-1+\epsilon}
\int_0^1 dx {x^{-1+\epsilon}\over (1-y+xy)^{-2+2\epsilon}}\\  
X_2 &=& 2\int_0^1 dy y^{-\epsilon}(1-y)^{-1+\epsilon}
\int_0^1 dx {x^{-1+\epsilon}\over (1-y+xy)^{-1+2\epsilon}} 
\ea
Let us first evaluate $X_1$. Performing a integration by parts with respect
to $x$ we obtain,
\ba
X_1 &=& \int_0^1 dy y^{-\epsilon}(1-y)^{-1+\epsilon}
\biggl[{1\over \epsilon} 
- {2(1-\epsilon)\over \epsilon(1+\epsilon)}y\nonumber\\
& &+ {2(1-\epsilon)(1-2\epsilon)\over \epsilon(1+\epsilon)}y^2
\int_0^1 dx x^{1+\epsilon}(1-y+xy)^{-2\epsilon}\biggr].
\ea
There remains a integration over $x$ within the third bracket. This
integrand is finite at $\epsilon=0$. So we first expand the integrand about
$\epsilon=0$ and then carry out the integration over $x$.
\ba
X_1 &=& \int_0^1 dy y^{-\epsilon}(1-y)^{-1+\epsilon}
\biggl[{1\over \epsilon} 
- {2(1-\epsilon)\over \epsilon(1+\epsilon)}y
+ {2(1-\epsilon)(1-2\epsilon)\over \epsilon(1+\epsilon)}
\Bigl\{ {1\over 2}y^2 + {5\over 4}\epsilon y^2 - \epsilon y\nonumber\\
& & - \epsilon (1-y)^2\ln(1-y)
+ {21\over 8}\epsilon^2 y^2
- {5\over 2}\epsilon^2 y 
+ \cdots + 0(\epsilon^3)\Bigr\}\biggr].
\ea 
Here dots imply the terms of order $\epsilon^2$, which after integration
over $y$ will contribute to order $\epsilon$ in $X_1$. Since we are
using minimal subtraction scheme, those terms of order $\epsilon$ in 
$X_1$ are not relevant for us. Carrying out the integration
over $y$ and retaining terms upto zeroth order in $\epsilon$ we 
obtain
\be
X_1 = {1\over \epsilon} + {3\over 2}.
\ee
In a similar fashion we can carry out the integration in $X_2$ and
the result is
\be
X_2 = {4\over \epsilon} + 4.
\ee
Using the result of $X_1$ and $X_2$ we can obtain $J_A^{(0)}$.
 
Using eqn.(\ref{fd phase space}) we perform the integration 
over $k$ and the result is
\be
J_A^{(1)} = {3m_A^2\over 16\pi^2\epsilon}
\left({m_A\over 4\pi\Lambda^2}\right)^{-\epsilon}
\left[\mu_A\sqrt{\mu_A^2 - m_A^2} 
- m_A^2\ln\left({\mu_A + \sqrt{\mu_A^2 - m_A^2}\over m_A}\right)\right].
\ee

\newpage

\begin{center}
{\bf Figure Captions}
\end{center}

\vspace{0.8cm}

Figure 1 : (a) Self energy diagram of quark at order $\alpha_s$.
(b) The counterterm diagram of quark two-point function at 
order $\alpha_s$.

Figure 2: (a) Gluon polarisation diagram at order $\alpha_s$
with only the quarks inside the loop.
(b) The counterterm diagram of gluon two-point function at 
order $\alpha_s$.

Figure 3: Diagrams contributing to the free energy density at order
$\alpha_s$. Here solid lines represent the quarks, coiled lines
represent the gluons and dashed lines represent the ghosts.

\newpage


\begin{center}
\begin{picture}(450,130)(0,0)
\GlueArc(110,80)(40,0,180){5}{8}
\GlueArc(110,80)(40,180,360){5}{8}
\Line(10,80)(70,80)
\Line(210,80)(150,80)
\Text(110,15)[]{$(a)$}
\Line(240,80)(300,80)
\Line(300,80)(360,80)
\Line(297,84)(303,76)
\Line(303,84)(297,76)
\Text(300,15)[]{$(b)$}

\end{picture}\\

Figure 1.

\end{center}



\begin{center}
\begin{picture}(450,130)(0,0)
\ArrowArc(110,80)(40,0,180)
\ArrowArc(110,80)(40,180,360)
\Gluon(10,80)(70,80){5}{4}
\Gluon(150,80)(210,80){5}{4}
\Text(110,15)[]{$(a)$}
\Gluon(240,80)(300,80){5}{4}
\Gluon(300,80)(360,80){5}{4}
\Vertex(300,80){5}
\Text(300,15)[]{$(b)$}

\end{picture}\\

Figure 2.

\end{center}


\newpage


\begin{center}
\begin{picture}(450,250)(0,0)
\ArrowArcn(110,200)(40,0,180)
\ArrowArcn(110,200)(40,180,360)
\Gluon(70,200)(150,200){5}{5}
\Text(110,135)[]{$(a)$}
\GlueArc(340,200)(40,0,180){5}{8}
\GlueArc(340,200)(40,180,360){5}{8}
\Gluon(300,200)(380,200){5}{5}
\Text(340,135)[]{$(b)$}
\DashArrowArcn(110,80)(40,0,180){3}
\DashArrowArcn(110,80)(40,180,360){3}
\Gluon(70,80)(150,80){5}{5}
\Text(110,15)[]{$(c)$}
\GlueArc(300,80)(35,-180,180){5}{16}
\GlueArc(380,80)(35,0,360){5}{16}
\Text(340,15)[]{$(d)$}
\ArrowArc(110,-40)(40,0,360)
\Line(107,-76)(113,-84)
\Line(107,-84)(113,-76)
\Text(110,-105)[]{$(e)$}
\GlueArc(340,-40)(35,-90,270){5}{16}
\Vertex(340,-80){5}
\Text(340,-105)[]{$(f)$}

\end{picture}\\

\vspace{7cm}

Figure 3.

\end{center}


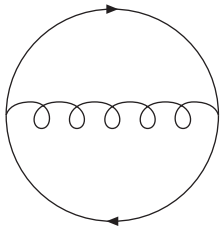
\end{document}